\documentclass[10pt,letterpaper,twocolumn]{article}
\usepackage{ol2}
\usepackage[draft]{hyperref}
\usepackage{amsmath}

\begin{document}

\twocolumn[

\title{Parallel real-time quantum random number generator}

\author{Xiaomin Guo$^{1,2}$, Chen Cheng$^{1,2}$, Mingchuan Wu$^{1,2}$, Qinzhong Gao$^{1,2}$, Pu Li$^{1,2}$, and Yanqiang Guo$^{1,2,*}$}

\address{$^{1}$Key Laboratory of Advanced Transducers and Intelligent Control System, Ministry of Education, Taiyuan University of Technology, Taiyuan 030024, China \\
$^{2}$College of Physics and Optoelectronics, Taiyuan University of Technology, Taiyuan 030024, China \\
*Corresponding author: guoyanqiang@tyut.edu.cn}

\begin{abstract}
Quantum random number generation exploits inherent randomness of quantum mechanical processes and measurements. Real-time generation rate of quantum random numbers is usually limited by electronic bandwidth and data processing rates. Here we use a multiplexing scheme to create a fast real-time quantum random number generator based on continuous variable vacuum fluctuations. Multiple sideband frequency modes of a quantum vacuum state within a homodyne detection bandwidth are concurrently extracted as the randomness source. Parallel post-processing of raw data from three sub-entropy sources is realized in one field-programmable gate array (FPGA) based on Toeplitz-hashing extractors. A cumulative generation rate of 8.25 Gbps in real-time is achieved. The system relies on optoelectronic components and circuits that could be integrated in a compact, economical package.
\end{abstract}

 ]

\noindent True random numbers are most critically required in cryptography for information security and communication network. The inherent randomness at the core of quantum mechanics makes quantum systems perfect entropy sources \cite{Acin16,Calude10}. A typical quantum random number generator (QRNG) prepares a well-defined quantum state, whose intrinsic quantum fluctuation is measured based on corresponding projective measurements close to theoretical idealization \cite{Svozil09,Collantes17}. In reality, devices inevitably induce classical noise so that the quantum randomness in the output is generally mixed with side information \cite{Gisin02,Bouda12,Li15}. Quantum randomness contained in raw data should be well quantified and extracted by applying informational theory provable randomness extractor (RE) \cite{Ma13,Lunghi15}.

In the past two decades, tremendous efforts have been made for responsible and practical QRNGs based on various quantum entropy sources \cite{Mitchell15,Ma16}. Vacuum state-based QRNG has become a promising quantum random number generation scheme \cite{Gabriel10,Haw15,Shi16}. This scheme is appealing for its convenience of state preparation, insensitivity of detection efficiency, much higher measurement bandwidth versus schemes based on qubit states \cite{Wei09,Fiorentino07,Wahl11} and more compact optical setup relative to phase noise measurements \cite{Guo10,Qi10}. Chip-size integration of this type QRNG is expectable because all components involved have been integrated on a single chip recently \cite{Abellan16,Raffaelli18,Lenzini18}. Furthermore, the theoretical model for the relationship between ideal random number exploiting process and implementation imperfections can be established explicitly, which provides basis for information theoretically secure RE \cite{Ma13}.

However, generic to all other QRNGs, serial type scheme and post-processing bottlenecks seriously restrict real-time generation rate of vacuum state-based QRNG. For applications in, such as, quantum key distribution or quantum digital signatures, random numbers with generation rate up to Gbps in real-time is consumed and even higher rate in the near future is required. Recently, real-time hash REs are realized based on field programmable gate array (FPGA) \cite{Zhang16,Zheng19}. It should be noted that high generation rates of some reported QRNG schemes even reach up to 10 Gbps, but they are theoretically equivalent results actually. Normally, the photodetector outputs were digitized and stored in a high-performance oscilloscope and then post-processed offline in a computer. While can be referred to evaluate the potential of the QRNG, these setups are bulky, fragile and expensive, and cannot be employed in practical real-time applications.

In this paper, different from previous reports, a multiplexed QRNG based on wideband vacuum noise is proposed as a solution to bottlenecks in the real-time rate. We present an optoelectronic system for simultaneously generating parallel, independent random sequences springing from quantum quadrature fluctuations of spectrally separated sideband frequency modes. Utilizing a single laser, a single homodyne detection system, and three nonoverlapping frequency sideband extractors, independent quantum random bit streams in parallel three paths are produced with a cumulative generation rate of 8.25 Gbps in real-time.

In terms of the establishment of the entropy source in frequency and time domain, we explore the limiting factors in the real-time rate of vacuum state-based QRNG.

\begin{figure*}[htbp]
\centering
\includegraphics[width=0.95\linewidth]{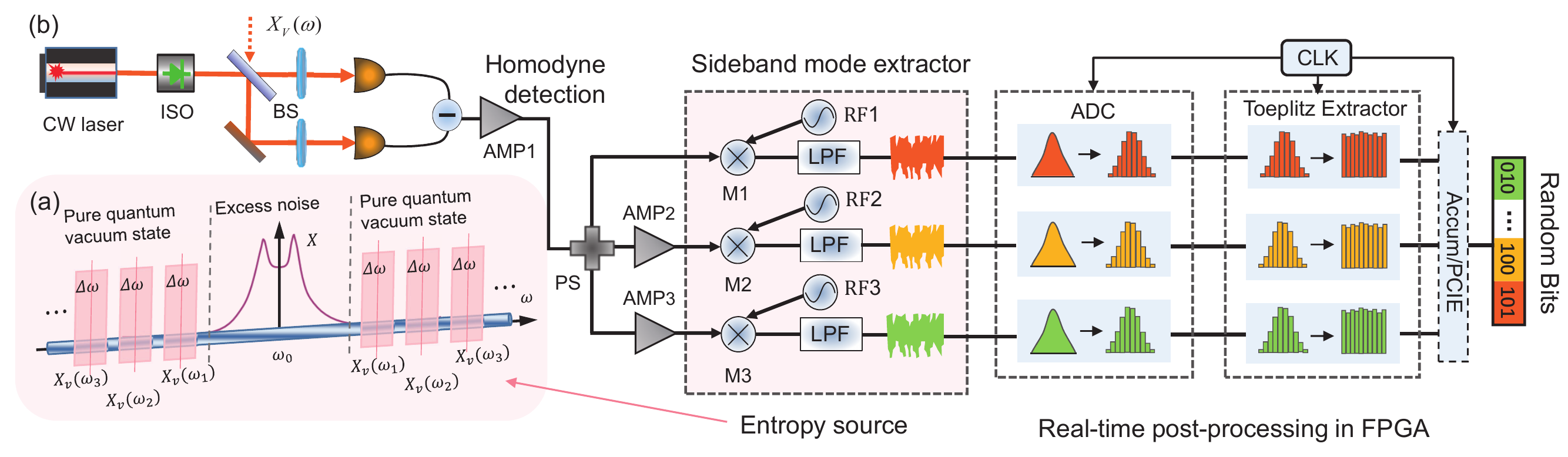}
\caption{(a) Schematic diagram of ingredients $X_{v}(\omega _{1}), X_{v}(\omega _{2}), X_{v}(\omega _{3})$ in random entropy source in our proposal. Each sub-entropy source is a composite quantum state consisting of a pair of adjacent sideband frequency modes relative to the optical carrier $X_{v}(\omega _{i})$. (b) Experimental scheme for the parallel real-time QRNG based on vacuum state measurement. CW, continuous wave; ISO, isolator; BS, 50/50 beamsplitter; PS, power splitter; AMP, ac amplifier; M, mixer; LPF, low pass filter; RF, radio-frequency generator; ADC, analog to digital converter; CLK, clock.}
\label{fig1}
\end{figure*}

In this scheme, quadrature fluctuations of continuous variable (CV) vacuum state are the source of randomness. Quadrature fluctuations of vacuum state induce ultra-wideband and uncontrollable white noise. Quantum modes at different frequencies are mutual independent and fluctuate following Gaussian distribution statistically \cite{Gabriel10,Cerf07}. Balanced homodyne detection (BHD) is employed to measure the quadrature fluctuations of vacuum state and amplifies them to macro level via local oscillator (LO) and electrical gains. Due to mechanical instabilities, relaxation oscillation and other detrimental effects in the LO, noise power spectrum (NPS) of the homodyne measurements is normally quite noisy at low frequency. Only above these frequencies, the laser modes are accommodated with pure vacuum state, as depicted in Fig. \ref{fig1}(a). Additionally, BHD gain cannot be flatted enough spanning the relatively large homodyne detection bandwidth attributed to limited gain-bandwidth product (GBWP) of op-amp. Nevertheless, accurate quantum entropy assessment of raw data heavily relies on signal-to-noise ratio (SNR) of the homodyne detection. For these reasons, in the frequency domain, quantum sideband mode at relatively high frequency with MHz bandwidth was usually chosen as genuine quantum entropy sources \cite{Gabriel10,Haw15,Shi16,Zhu12} for such kind of QRNG.

In time domain, the entropy source is defined upon the frequency mode bandwidth. That is, in fact if the bandwidth is 1 MHz, a new quantum state arrives every 1 $\mu s$ and the measurement outcome corresponds to the eigenvalue of the observable averaged among 1 $\mu s$. So the upper limit for random bit generation rate ultimately depends on the frequency mode bandwidth according to Nyquist rule:
\begin{equation}
C_{\max }=P_{H}\times n\times 2W_{BW}.
\label{eq1}
\end{equation}
Here $P_{H}=H_{\min }(X\mid E)/n$ expresses the quantum conditional min-entropy per bit, $n$ is the digital resolution, and $W_{BW}$ is the bandwidth of the sideband frequency mode. For real-time QRNG, the RE should process such amount of raw data at once.

As a result, not the bandwidth of homodyne detector but of the side-mode quantum entropy source, as well as limited post-processing power restrict the real-time generation speed of the vacuum-based QRNG. In fact, quantum entropy source within the homodyne bandwidth is often not fully exploited.

Based on above considerations, we propose concurrently extract multiple nonoverlapping sideband quantum states within the BHD bandwidth and realize post-processing in parallel in one FPGA. We construct the entropy source as schemed in Fig. \ref{fig1}(a). Sideband modes with distinct center frequencies are chosen relative to the carrier. Appropriate bandwidth of these modes is decided based on relatively constant SNR. Quantum conditional min-entropy is rigorously and independently evaluated for each sub-entropy source, that is, each quantum sideband frequency mode at different frequency. Real-time post-processing of raw data from the three paths are realized based on parallel algorithm advantage of FPGA.

In Fig. \ref{fig1}(b), we show the experimental scheme of the vacuum state-based parallel QRNG. A 1550 nm laser diode (LD) driven by constant current provides the LO. Single-mode CW beam with a power of 4.8 mW is incident into one port of a 50/50 beamsplitter, while the other input port is blocked to ensure that only the vacuum state enters in. Interfered vacuum and LO fields are output with balanced power and detected by BHD (PDB480C, Thorlabs Inc., USA). Common-mode classical noise in laser is cancelled and the quadrature of the vacuum state is amplified.

\begin{figure}[htbp]
\centering
\includegraphics[width=0.75\linewidth]{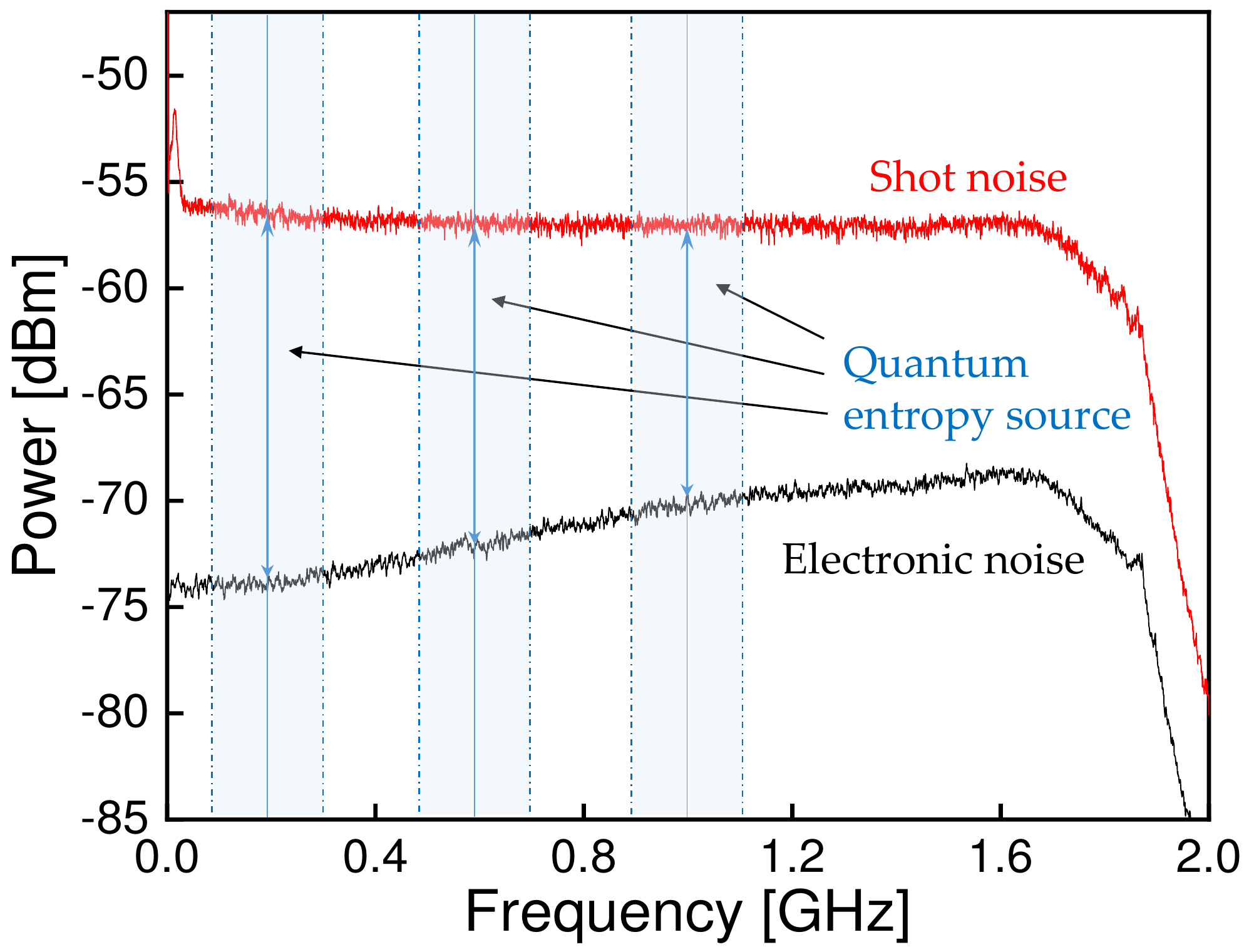}
\caption{Amplified vacuum noise power spectrum when LO power is 4.8 mW. 120 MHz sideband modes center at 200 MHz, 600 MHz, and 1 GHz are filtered out as entropy source of the parallel QRNG.}
\label{fig2}
\end{figure}

Firstly, NPS of BHD measurements is recorded by a spectral analyzer (N9010A, Agilent Technologies Inc., USA). As shown in Fig. \ref{fig2}, classical noise in the photocurrents is rejected effectively over the whole detection band with SNR above 10 dB. However, the clearance has evident dependence on analysis frequency as discussed above, which induces nonuniform quantum classical noise ratio (QCNR) and hinders conformance assessment of quantum min-entropy within the detection bandwidth. Based on the moderate processing power of our FPGA (xc7k325t, Xilinx Inc., USA), we extract three quantum sideband frequency modes with bandwidths of 120 MHz within the measurement bandwidth. The signal outcome from the BHD is divided into three parts via a power splitter. The first component is mixed down with a 200 MHz rf signal and then passes through a low-pass-filter (LPF) of 120 MHz cut-off frequency. In this way, the sub-entropy source is defined as a composite quantum state centered around 200 MHz relative to the optical carrier with a bandwidth of 120 MHz. The second and the third parts are mixed down with 600 MHz and 1 GHz rf signals independently and filtered with 120 MHz LPFs as well. The three sideband frequency modes of the vacuum state work together to contribute quantum randomness to the QRNG. 16-bit analog to digital converter (ADC) are used to transform the analog signals from each path into binary random sequences. Sampling clock of every ADC is set to 240 MHz, upper limit of twice the LPF bandwidth, to avoid temporal correlations between samples.

\begin{figure}[htbp]
\centering
\includegraphics[width=\linewidth]{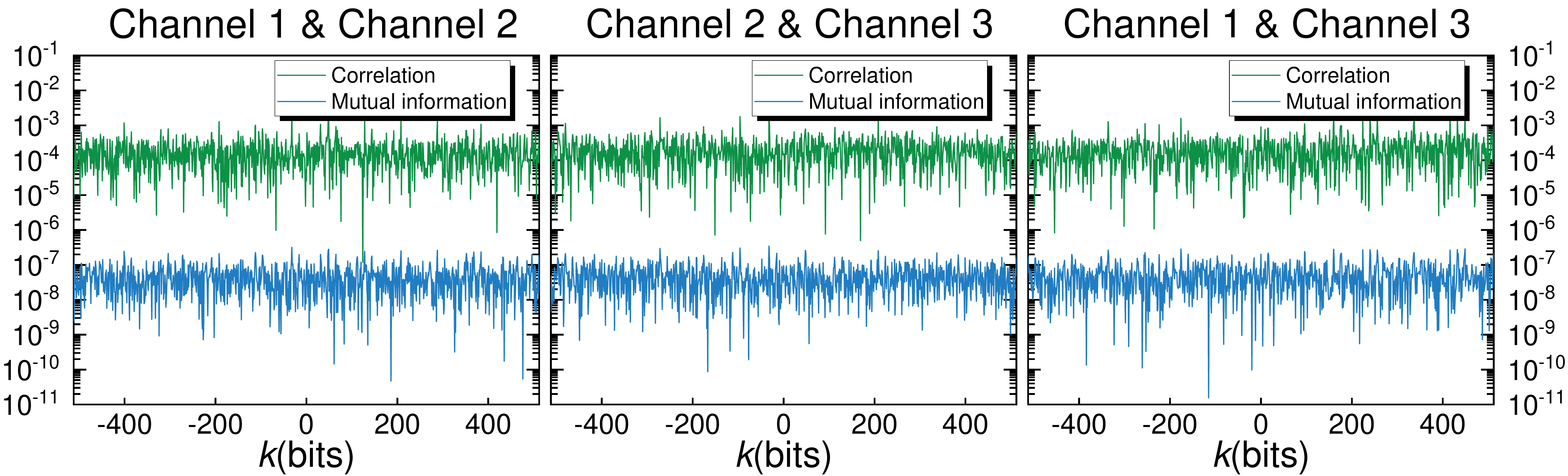}
\caption{Statistically computed absolute correlation $\rho _{xy}[k]$ and mutual information $I _{xy}[k]$ between two $10^{9}$ bit sequences of channel 1 and channel 2 (a), channel 2 and channel 3 (b), channel 1 and channel 3 (c).}
\label{fig3}
\end{figure}

For testifying independence between the channels, correlations of every two binary sequences are statistically computed. Furthermore, we calculated the mutual information of them, $I_{ab}$, as shown in Fig. \ref{fig3}. The parallel binary sequences exhibit low bias and undetectable interchannel correlation \cite{Cover91}.

\begin{figure}[htbp]
\centering
\includegraphics[width=0.8\linewidth]{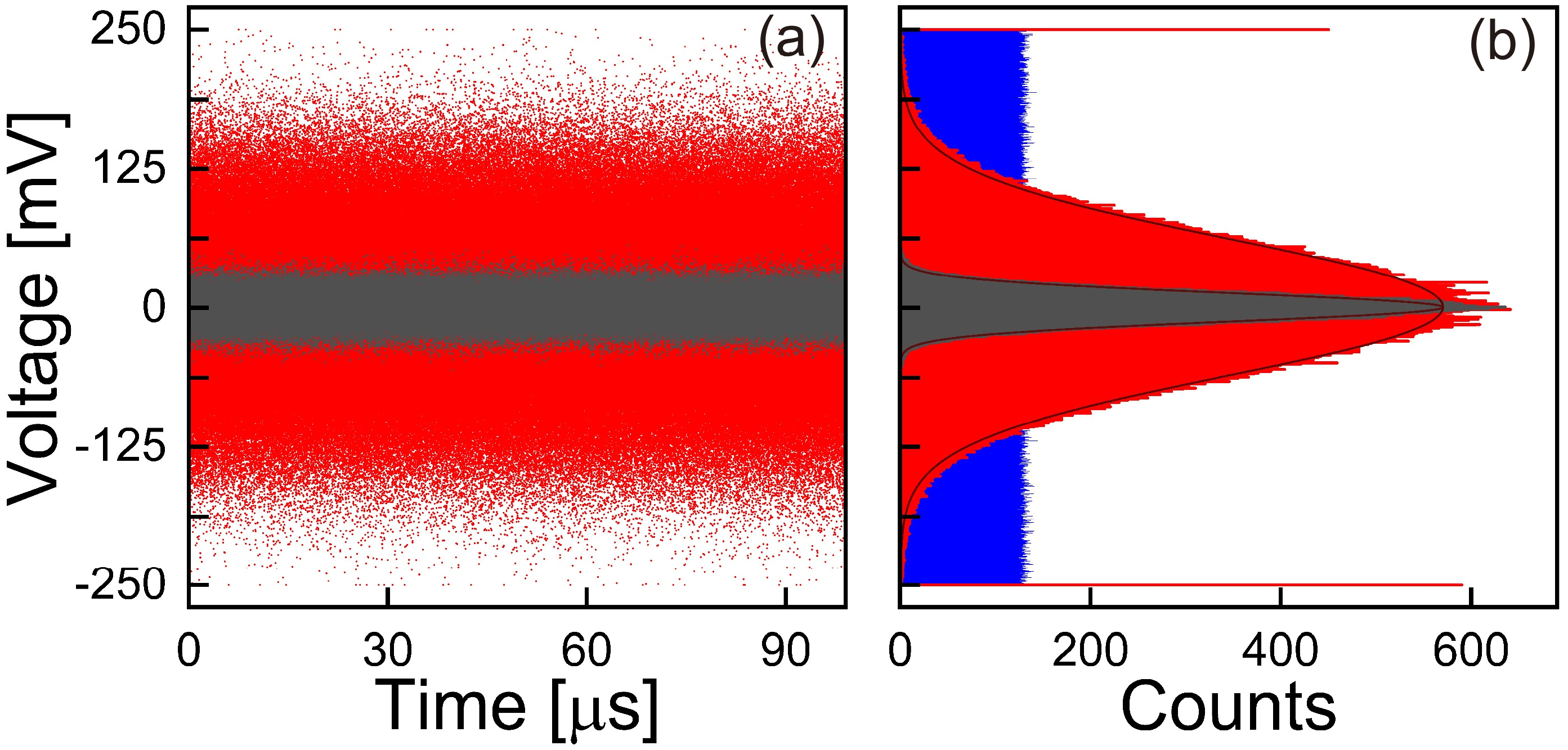}
\caption{(a) Typical 10 million raw data collected under optimal ADC sampling range from the sub-entropy source at 200 MHz; (b) Histogram of the vacuum (red), electronic noise (gray) and random numbers extracted from the raw data based on Toeplitz-hashing extractor (blue).}
\label{fig4}
\end{figure}

Using a representative sample of 10 Mbits per channel, we calculated the worst-case min-entropy $H_{\min }(X\mid E)$ conditioned on best sampling range. Entropy sourcing from the measurement of the vacuum state is evaluated separately and rigorously for each sub-entropy source. By choosing the analog-digital conversion range appropriately and tuning the LO intensity finely, the amount of off-scale points of path 1 is controlled within allowed statistical deviation, as shown in Fig. \ref{fig4}. While due to their relatively low QCNR, a few unused bins exist and induce too many blocks of zeros and ones in raw data from sideband modes 2 and 3. A low bandwidth and gain AC amplifier is inserted into each of the two paths to adjust the signal intensity independently. Min-entropy of raw data from each path is calculated as 14.2, 13.5, and 12.9 respectively \cite{Guo18}.

Toeplitz-hashing extractor, one of the most widely acknowledged information-theoretically provable REs, is employed in post-processing phase. The amount of quantum randomness extractable from the raw data is evaluated by the leftover hash lemma based on the conditional min-entropy \cite{Tomamichel11}
\begin{equation}
\ell \leq N\times H_{\min }(X\mid E)-\log \frac{1}{\varepsilon ^{2}}.
\label{eq2}
\end{equation}
Here $N$ is the number of samples and $\varepsilon $ is the hash security parameter which represents the distance between the string produced by the randomness extractor and a perfectly uniform random string.

Toeplitz REs for the three quantum modes are implemented in real-time in one FPGA. FPGA's concurrent processing character makes it very suitable for post-processing of our multiplexed random number generation. On one hand, Toeplitz matrix algorithm can be realized in high speed since multiplication and addition between different columns are completely independent. On the other hand, raw data extracted from each vacuum sideband mode can be independently post-processed concurrently in the FPGA with high resource-efficient. Due to the limitation of FPGA hardware resource, it is impossible to directly run a large matrix rapidly. In our work, we construct appropriate matrixes to match the scale of the FPGA and achieved high extraction ratio without extra reduction in entropy \cite{Zhang16}. Furthermore, for each quantum sideband mode, the Toeplitz RE is realized in a concurrent pipeline algorithm of three modules, including matrix building from seeds, submatrix multiplication and vector accumulation in a register. Finally, random numbers are transmitted to a computer in real-time. Based on FPGA's internal clock of 240 MHz, ADC acquisition of three quantum modes, three modules in each pipeline and PCI-E transmission are serviced for synchronization.

\begin{figure}[htbp]
\centering
\includegraphics[width=0.9\linewidth]{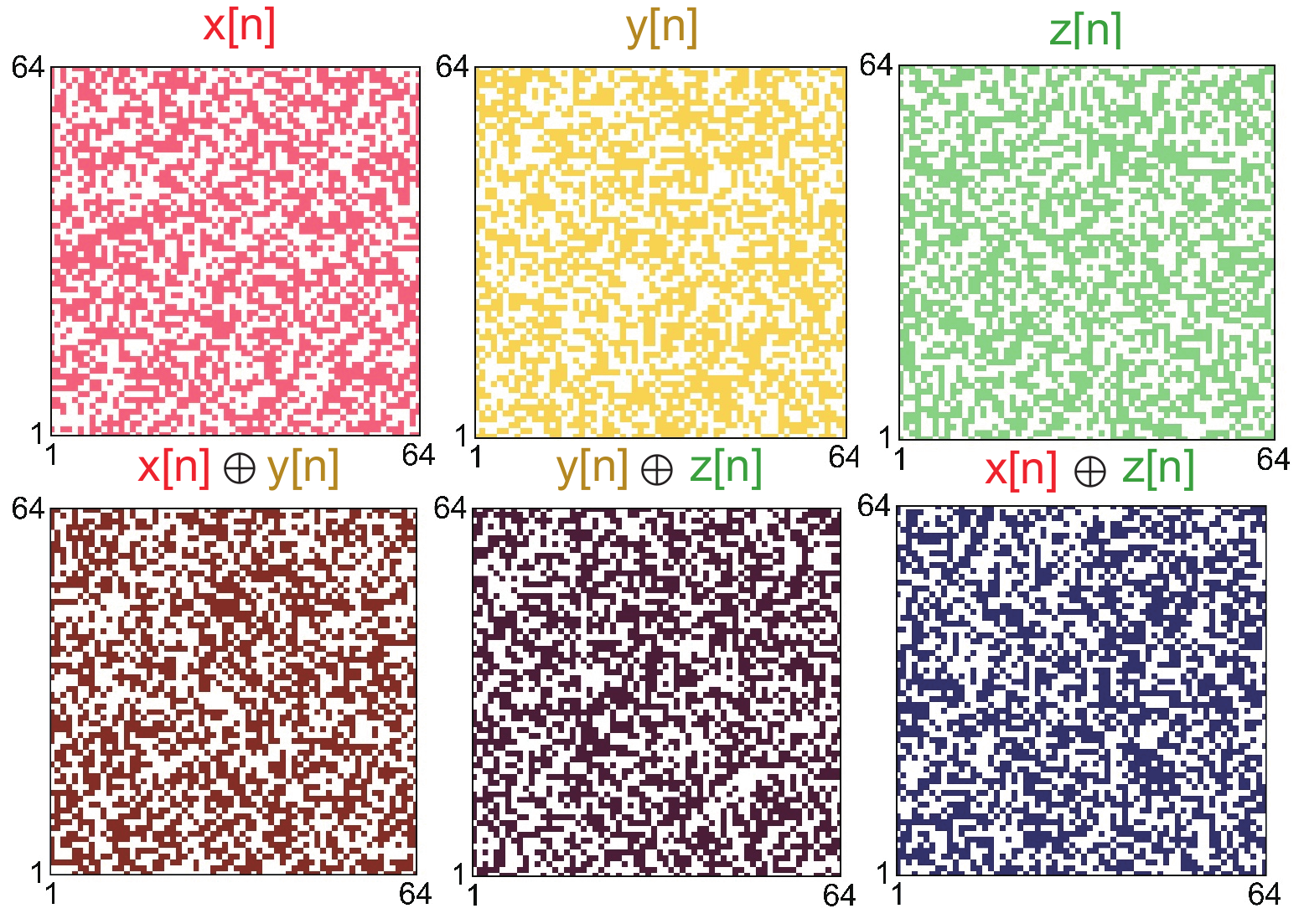}
\caption{The $64\times 64$ random bitmap images from three channels random bit streams (a, b, c). The (d, e, f) panels show the XOR interchannel.}
\label{fig5}
\end{figure}

For quantum mode centering at 200 MHz and the min-entropy of 14.2, extraction ratio of 75.7\% is calculated based on Eq. (\ref{eq2}) with security parameter of $2^{-50}$ and the matrix scale is set as $581\times 768$. Based on the min-entropy of 13.5 and 12.9 of quantum modes center at 600 MHz and 1 GHz, matrixes of $548\times 768$ and $519\times 768$ are established with the same security parameter. Real-time generation rates for the three channels are 2.91 Gbps, 2.74 Gbps, and 2.60 Gbps respectively. As a result, we produce random numbers in real-time with a cumulative rate of 8.25 Gbps. In this process, calculation resources of the FPGA are scaling with the path number, and the matrix operation of the three Toeplitz REs consumes 43.8\% of the FPGA internal resource. While other function modules, including the I/O, take a partition of 30\%$\sim $35\% of the FPGA internal resource consistently. In this way, the hardware resource of FPGA is applied more efficiently.

For testifying randomness of random number stream extracted from each sub-entropy source and independence between them, we construct $64\times 64$ random bitmap images of the three binary sequences and computed XOR between every two of them. Respective bitmap images from all three data channels exhibit no apparent pattern or bias. XOR bitmaps show no correlation with their original sequences, as shown in Fig. \ref{fig5}.

\begin{figure}[htbp]
\centering
\includegraphics[width=0.95\linewidth]{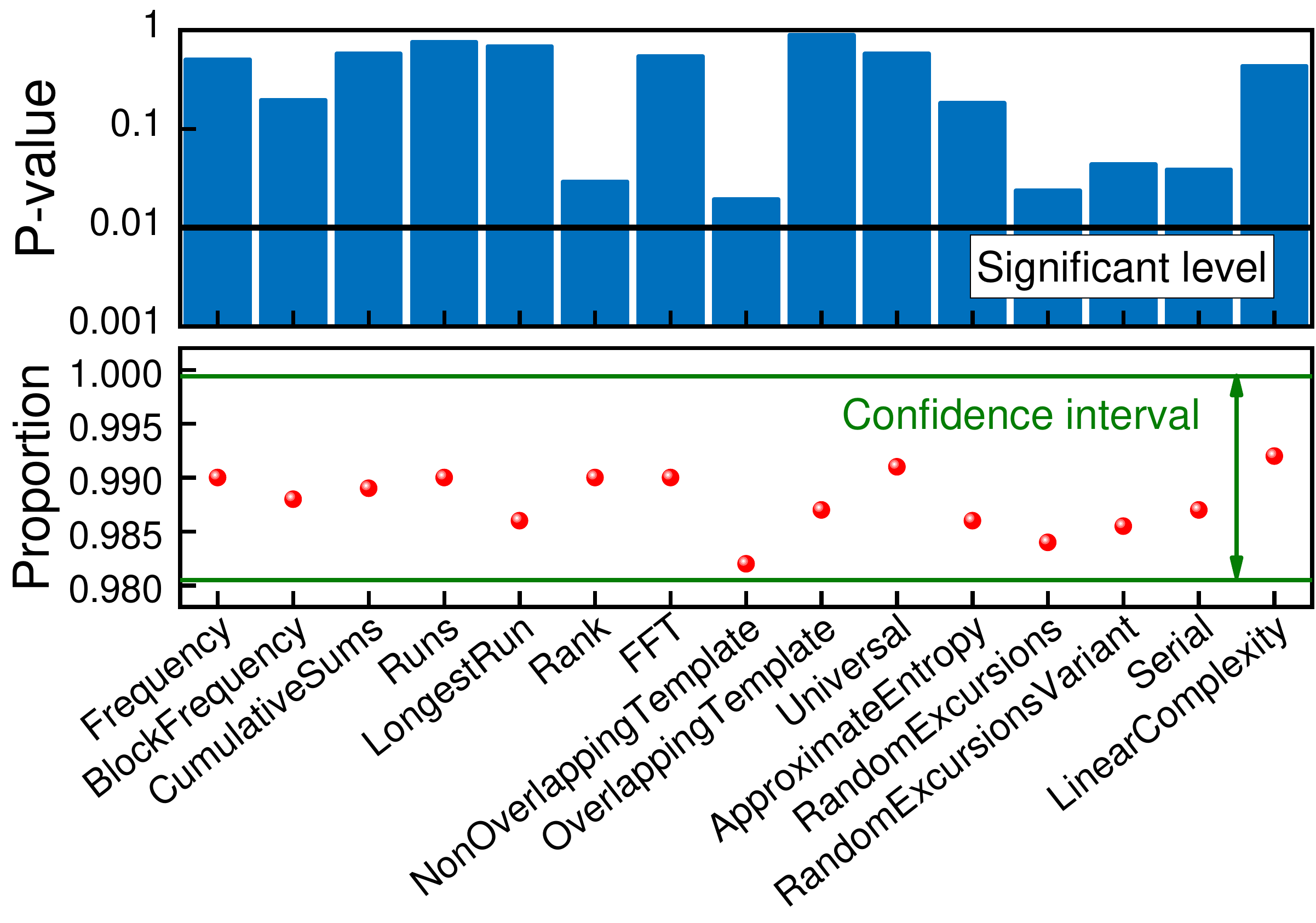}
\caption{Typical results of standard NIST Statistical Test Suite. For each test term, the bar values represent the P values of the worst cases of our test outcomes. The red dots indicate that the minimum pass rate for each statistical test is 98.2\%, which certifies a good level of true randomness.}
\label{fig6}
\end{figure}

At last, true random numbers from three paths are cumulated and output to PC via PCI-E. 1000 samples with each 1 million bits are subjected to NIST statistical test suite \cite{NIST} and the significant level is set as $\alpha =0.01$. The test is successful if final P-values of all samples are higher than $\alpha $ with a proportion within the confidence interval of $(1-\alpha )\pm 3\sqrt{(1-\alpha )\alpha /n}=0.99\pm 0.00944$ for all 15 test suits. The test report is shown in Fig. \ref{fig6}. The final data successfully pass all NIST tests.

In summary, we report an extensible high-speed, real-time QRNG based on CV quantum vacuum state. Taking advantage of the full bandwidth of our BHD, real-time generation rate above 34.8 Gbps of quantum random numbers will be accessible. The scheme combines the wide band properties of quantum vacuum noise and homodyne detector, and take the parallel algorithm advantage of FPGA. The system uses only commercially available optoelectronic components that could be integrated at the chip- or board-level. The work provides an extensible and cost-efficient method for ultrafast parallel QRNG and can substantially push the practicality of high-performance RNGs in cryptography applications.

\vspace{0.15in}
\noindent\textbf{Funding.} National Natural Science Foundation of China (NSFC) (61875147, 61731014, 61671316, 61775158); Shanxi Scholarship Council of China (SXSCC) (2017-040); Natural Science Foundation of Shanxi Province (201701D221116, 201801D221182); Scientific and Technological Innovation Programs of Higher Education Institutions in Shanxi (STIP) (201802053, 2019L0131).

\vspace{0.15in}
\noindent\textbf{Acknowledgment.} We thank Prof. Tiancai Zhang from State Key Laboratory of Quantum Optics and Quantum Optics Devices, Institute of Opto-Electronics, Shanxi University, China, for his very useful comments and advice.

% Bibliography
%\bibliography{sample}

\begin{thebibliography}{99}
\bibitem{Acin16} A. Ac\'{\i}n and L. Masanes, Nature \textbf{540}, 213 (2016).

\bibitem{Calude10} C. S. Calude, M. J. Dinneen, M. Dumitrescu, and K. Svozil, Phys. Rev. A \textbf{82}, 022102 (2010).

\bibitem{Svozil09} K. Svozil, Phys. Rev. A \textbf{79}, 054306 (2009).

\bibitem{Collantes17} M. Herrero-Collantes and J. C. Garcia-Escartin, Rev. Mod. Phys. \textbf{89}, 015004 (2017).

\bibitem{Gisin02} N. Gisin, G. Ribordy, W. Tittel, and H. Zbinden, Rev. Mod. Phys. \textbf{74}, 145 (2002).

\bibitem{Bouda12} J. Bouda, M. Pivoluska, M. Plesch, and C. Wilmott, Phys. Rev. A \textbf{86}, 062308 (2012).

\bibitem{Li15} H. W. Li, Z. Q. Yin, S. Wang, Y. J. Qian, W. Chen, G. C. Guo, and Z. F. Han, Sci. Rep. \textbf{5}, 16200 (2015).

\bibitem{Ma13} X. F. Ma, F. H. Xu, H. Xu, X. Q. Tan, B. Qi, and X. K. Lo, Phys. Rev. A \textbf{87}, 062327 (2013).

\bibitem{Lunghi15} T. Lunghi, J. Brask, C. Lim, Q. Lavigne, J. Bowles, A. Martin, H. Zbinden, and N. Brunner, Phys. Rev. Lett. \textbf{114}, 150501 (2015).

\bibitem{Mitchell15} M. W. Mitchell, C. Abellan, and W. Amaya, Phys. Rev. A \textbf{91}, 012314 (2015).

\bibitem{Ma16} X. F. Ma, Y. Xiao, C. Zhu, Q. Bing, and Z. Zhen, npj Quantum Inform \textbf{2}, 16021 (2016).

\bibitem{Gabriel10} C. Gabriel, C. Wittmann, D. Sych, R. Dong, W. Mauerer, U. L. Andersen, C. Marquardt, and G. Leuchs, Nat. Photonics \textbf{4}, 711 (2010).

\bibitem{Haw15} J. Y. Haw, S. M. Assad, A. M. Lance, N. H. Y. Ng, V. Sharma, P. K. Lam, and T. Symul, Phys. Rev. Applied \textbf{3}, 054004 (2015).

\bibitem{Shi16} Y. C. Shi, B. Chng, and C. Kurtsiefer, Appl. Phys. Lett. \textbf{109}, 041101 (2016).

\bibitem{Wei09} W. Wei and H. Guo, Opt. Lett. \textbf{34}, 1876 (2009).

\bibitem{Fiorentino07} M. Fiorentino, C. Santori, S. M. Spillane, R. G. Beausoleil, and W. J. Munro, Phys. Rev. A \textbf{75}, 032334 (2007).

\bibitem{Wahl11} M. Wahl, M. Leifgen, M. Berlin, T. Rohlicke, H. J. Rahn, and O. Benson, Appl. Phys. Lett. \textbf{98}, 171105 (2011).

\bibitem{Guo10} H. Guo, W. Tang, Y. Liu, and W. Wei, Phys. Rev. E \textbf{81}, 051137 (2010).

\bibitem{Qi10} B. Qi, Y.-M. Chi, H.-K. Lo, and L. Qian, Opt. Lett. \textbf{35}, 312 (2010).

\bibitem{Abellan16} C. Abellan, W. Amaya, D. Domenech, P. Mu\~{n}oz, J. Capmany, S. Longhi, M. W. Mitchell, and V. Pruneri, Optica \textbf{3}, 989 (2016).

\bibitem{Raffaelli18} F. Raffaelli, G. Ferranti, D. H. Mahler, P. Sibson, J. E. Kennard, A. Santamato, G. Sinclair, D. Bonneau, M. G. Thompson, and J. C. F. Matthews, Quantum Sci. Technol. \textbf{3}, 025003 (2018).

\bibitem{Lenzini18} F. Lenzini, J. Janousek, O. Thearle, M. Villa, B. Haylock, S. Kasture, L. Cui, H. P. Phan, D. V. Dao, H. Yonezawa, P. K. Lam, E. H. Huntington, and M. Lobino, Sci. Adv. \textbf{4}, eaat9331 (2018).

\bibitem{Zhang16} X. G. Zhang, Y. Q. Nie, H. Zhou, H. Liang, X. Ma, J. Zhang, and J. W. Pan, Rev. Sci. Instrum. \textbf{87}, 076102 (2016).

\bibitem{Zheng19} Z. Zheng, Y. Zhang, W. Huang, S. Yu and H. Guo, Rev. Sci. Instrum. \textbf{90}, 043105 (2019).

\bibitem{Cerf07} N. J. Cerf, G. Leuchs, and E. S. Polzik, ''Quantum Information with Continuous Variables of Atoms and Light,'' (Imperial College Press, 2007).

\bibitem{Zhu12} Y. Y. Zhu, G. Q. He, and G. H. Zeng, Int. J. Quantum Inf. \textbf{10}, 1250012 (2012).

\bibitem{Cover91} T. M. Cover and J. A. Thomas, ''Elements of Information Theory,'' (Wiley-Interscience, 1991).

\bibitem{Guo18} X. M. Guo, R. P. Liu, P. Li, C. Cheng, M. C. Wu, and Y. Q. Guo, Entropy \textbf{20}, 819 (2018).

\bibitem{Tomamichel11} M. Tomamichel, C. Schaffner, A. Smith, and R. Renner, IEEE T. on Inform. Theory \textbf{57}, 5524 (2011).

\bibitem{NIST} http://csrc.nist.gov/groups/ST/toolkit/rng/index.html.
\end{thebibliography}

\end{document}